\documentclass{elsart}
\usepackage{epsfig}
\usepackage{amssymb}
\usepackage{amsmath}
\usepackage{colordvi}
\begin{document}
\begin{frontmatter}

\title{Spin-orbit coupling in bulk GaAs}

\author[nl,phys]{J. Y. Fu},
\author[phys]{M. Q. Weng},
\author[nl,phys]{M. W. Wu\thanksref{cor}}
\corauth[cor]{Corresponding author.
Telephone: +86-551-3603524;
Fax: +86-551-3603524.
}
\ead{mwwu@ustc.edu.cn.}
\address[nl]{Hefei National Laboratory for Physical Sciences at
  Microscale, University of Science and Technology of China,
  Hefei, Anhui, 230026, China}
\address[phys]{Department of Physics, University of Science and
  Technology of China, Hefei, Anhui, 230026, China
\thanksref{mailing}}
\thanks[mailing]{Mailing address.}

\begin{abstract}

We study the spin-orbit coupling in the whole Brillouin zone for
GaAs using both the $sp^3s^{\ast}d^5$ and $sp^3s^{\ast}$
nearest-neighbor tight-binding models. In the $\Gamma$-valley, the
spin splitting obtained is in good agreement with experimental data.
We then further explicitly present the coefficients of the spin
splitting in GaAs $L$ and $X$ valleys. These results are important
to the realization of spintronic device and the investigation of
spin dynamics far away from equilibrium.
\end{abstract}

\end{frontmatter}


\section{Introduction}

Spin-orbit coupling (SOC) is the key ingredient to the semiconductor
spintronic devices \cite{wolf,zutic_2004}.
Most of the proposed schemes of electrical
generation, manipulation and detection of electron spin rely on
it. Complete understanding of the SOC is therefore of great importance. In the
bulk zinc-blende-type semiconductor such as GaAs, it is well known
that near the center of Brillouin zone the zero field splitting caused
by the SOC depends cubically on the wave-vector $k$ due to the bulk inversion
asymmetry \cite{dp,dpb}
or linearly due to the structure inversion asymmetry \cite{ras,rashba}.
There are few investigations of the SOC for the states away from the
band edge. The
{\it ab initio} band structure calculation \cite{Cardona},
diagonalizing of truncated
$\mathbf{k}\cdot\mathbf{p}$ Hamiltonian \cite{Cardona}
or nearest-neighbor tight-binding (TB) model including the SOC
\cite{Jancu1,Jancu2} have been
performed to obtain the spin-orbit splitting outside of the Brillouin
zone center.
For the states near other high symmetry points such as
the $L$ and $X$ points, one can get the form of the splitting from the
symmetry property \cite{ivchenko}, whereas the actual
coefficients need to be further
calculated. Recently the splitting of the $L$-valley in bulk GaSb and
GaSb/AlSb quantum wells were calculated using an $sp^3s^\ast d^5$
nearest-neighbor TB model including the SOC \cite{Jancu1}.
It is shown that the splitting in GaSb $L$-valley exceeds
10~meV, an order of magnitude larger than the typical value in the
$\Gamma$-valley. For GaAs the corresponding data are still not
available.

The lack of quantitative information of the SOC outside the Brillouin
center is not crucial to the development of spintronics at the present
stage, since the electrons in most of the study locate at the bottom of the
$\Gamma$-valley. However, in real situation, the devices usually work
under high electric field which can drive the electrons to the states far
away from $\Gamma$ point, or even further to other valleys such as $L$-
and/or $X$-valleys. Therefore for the realization of the spintronic
devices, the SOC in the whole Brillouin zone is essential. In the
previous works on high field spin
transport in GaAs \cite{saikin_06,shen_2004}, the
coefficient of spin splitting in high valleys are approximated by
other material such as GaSb due to the lack of the corresponding data for
GaAs. In this report we present the spin splitting of GaAs for the
whole lowest conduction band.
Especially, we calculate the the coefficients of spin splitting
in $L$- and $X$- valleys.

\section{Calculation and Results}
Our calculations are performed in $sp^3s^*$ and $sp^3s^*d^5$
nearest-neighbor TB models with the SOC.
This method has been proven to be an effective approach in band
structure calculations \cite{Jancu2,Vogl,Santos,Klimeck,Boykin1,Boykin2,Diaz}.
The parameters we use are adopted from the published
literate \cite{Jancu2,Vogl,Santos,Klimeck,Boykin1,Boykin2,Diaz}.
The original parameter sets have been optimized to fit the experimental
data, such as the band edge and the effective mass. With the comparison to the
experimental data, the spin splitting near the $\Gamma$-point is
calculated as the benchmark of the suitability of these parameters for
calculating the spin splitting.

\begin{figure}[htbp]
  \centering
  \epsfig{file=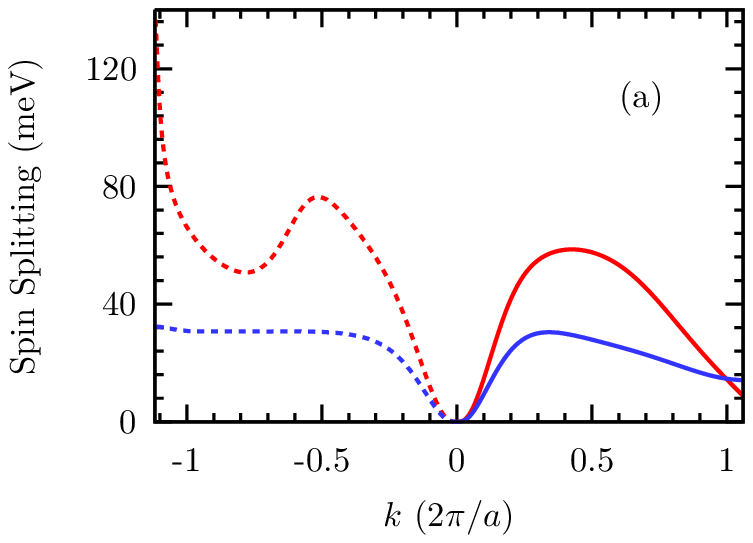,width=6cm}\\
\vskip1.2pc
  \epsfig{file=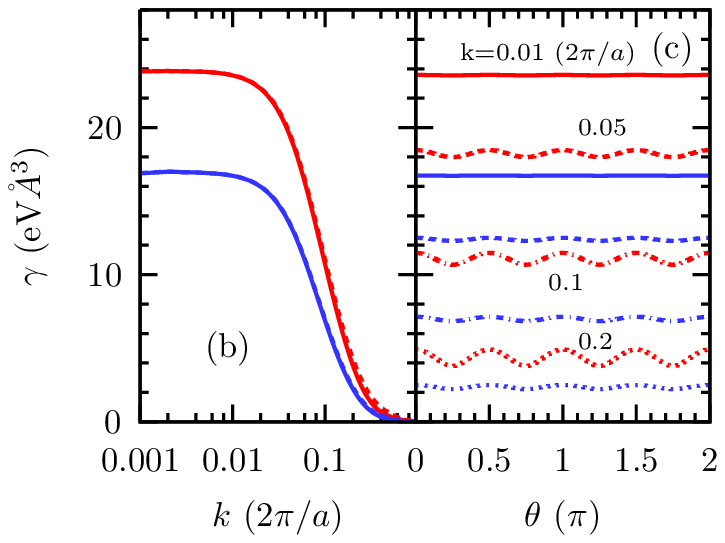,width=6cm}
  \caption{(Color Online) (a) Spin splitting in $\Gamma$-valley for
    different momentum
    along $\Gamma$-$K$ (solid curves) and $\Gamma$-$W$ (dashed curves)
    directions;
    (b) The corresponding SOC coefficient $\gamma$ {\it
      vs.} momentum along $\Gamma$-$K$ (solid curves) and $\Gamma$-$W$
    (dashed curves) directions;
    (c) $\gamma$ {\it vs.} angle of the momentum with different
    amplitudes. Solid curves: $k=0.01$; Dashed curves: $k=0.05$; Chain curves: $k=0.1$;
    Dotted curves: $k=0.2$\ $(2\pi/a)$, respectively.
    Red/Blue curves are the results from
    the $sp^3s^{\ast}d^5$/$sp^3s^{\ast}$ model.
  }
  \label{fig:gamma}
\end{figure}

In Fig.\ \ref{fig:gamma} we show the spin splitting around $\Gamma$
point for different momentums. In the figure the parameter sets are
chosen from Refs.\ \cite{Klimeck} and \cite{Jancu2} for
$sp^{3}s^{\ast}$ and $sp^3s^{\ast}d^5$ respectively.
One can see from the figure that
for small momentum the splittings calculated from $sp^3s^{\ast}$ and
$sp^3d^5s^{\ast}$ approaches both increase with the momentum. When the
momentum becomes large, the splitting is no longer a monotonic.
 The splittings along different directions have different
behaviors. For the states near the $\Gamma$-point, the splitting can be
described by 
$\gamma\mathbf{\Omega}(\mathbf{k})\cdot\mbox{\boldmath$\sigma$\unboldmath}$
with \mbox{\boldmath$\sigma$\unboldmath}
being the Pauli matrices.
In the coordinate system of $\hat{x}=[100]$, $\hat{y}=[010]$ and
$\hat{z}=[001]$,
\begin{equation}
  \label{eq:gamma}
\gamma  \mathbf{\Omega}(\mathbf{k})=\gamma
  (k_x(k_y^2-k_z^2),k_y(k_z^2-k_x^2),k_z(k_x^2-k_y^2)).
\end{equation}
  This gives rise to a splitting that varies cubically in $k$,
    i.e., $\Delta E\sim \gamma k^3$.
The splitting along different directions
are different. In the $x$-$y$ plane, the splitting is
$\Delta E=\gamma k^3|\sin (2\theta)|$ where $\theta$ is the angle
between the momentum and $(100)$-axis.
The exact value of $\gamma$ is still in debate.
Different approaches give various values range from 8.5 to
34.5\ eV$\cdot$\AA$^3$. Overviews of these results are nicely listed in
Refs.\ \cite{krich_07,chantis_06}.
Theoretical calculations based on different approaches give quite
different values. There are two kinds of experiments that can
measure $\gamma$ value. One is through the direct measurement of the
splitting using Raman scattering. Experiments based on this approach
show that $\gamma$ is about 23.5\ eV$\cdot$\AA$^3$ in wide GaAs
quantum well \cite{richards_93}. In asymmetric GaAs/AlGaAs
heterostructure/quantum well, $\gamma$ is about 16.5 or 11.0\
eV$\cdot$\AA$^3$ \cite{jusserand_95,richards_96}. The other kind of
measurement is through spin relaxation time or magneto-conductance.
This kind of measurement is indirect since it depends on how to
qualitatively calculate the spin relaxation time or
magneto-conductance. Earlier works of this kind estimate that
$\gamma$ is about 20-30\ eV$\cdot$\AA$^3$
\cite{maruschak_1983,aronov,miller_03}. Recent calculation based on
fully microscopic approach reveals that the experiments in two-dimensional (2D)
 system can  be explained by using much smaller $\gamma$ value \cite{zhou_prb_2007}. In
our calculation, value of this coefficient is calculated as
$\gamma(\mathbf{k})=\Delta E/(2|\Omega(\mathbf{k})|)$. The results
for different momentums are shown in Fig.\ \ref{fig:gamma}(b) and
(c). One can see from the figure that for $k<0.04\pi/a$, $\gamma$ is
almost a constant that is independent on magnitude and the angle of
momentum. For the parameters we use, $sp^3s^{\ast}$ and
$sp^3s^{\ast}d^5$ give $\gamma=17.0$ and $23.9$\ eV$\cdot$\AA$^3$
near the $\Gamma$-point, respectively. Both are close to the
experimental data from Raman scattering. This good agreement between
the theoretical result and the experimental data shows  that the
parameters we use can be applied to study the spin splitting for
whole Brillouin zone. One can see from the figure that both models
predict that the value of $\gamma$ decreases with the increase of
momentum\footnote{This explains the  small value obtained
in 2D system \cite{zhou_prb_2007}.}. Moreover, angle dependence of $\gamma$ becomes remarkable
for large momentum. Thus $\gamma$ is no longer a constant for large
momentum. Practically the SOC described by Eq.\ (\ref{eq:gamma})
with constant $\gamma$ is a good approximation for the state not far
away from equilibrium since even for high-carrier-density samples
the value of $\gamma$ at Fermi surface is only a few percents
smaller than the value at $k=0$. However, when  electrons are driven
far away from the $\Gamma$ point, Eq.\ (\ref{eq:gamma}) is expected
to over-estimate the SOC.

\begin{figure}[htbp]
  \centering
  \epsfig{file=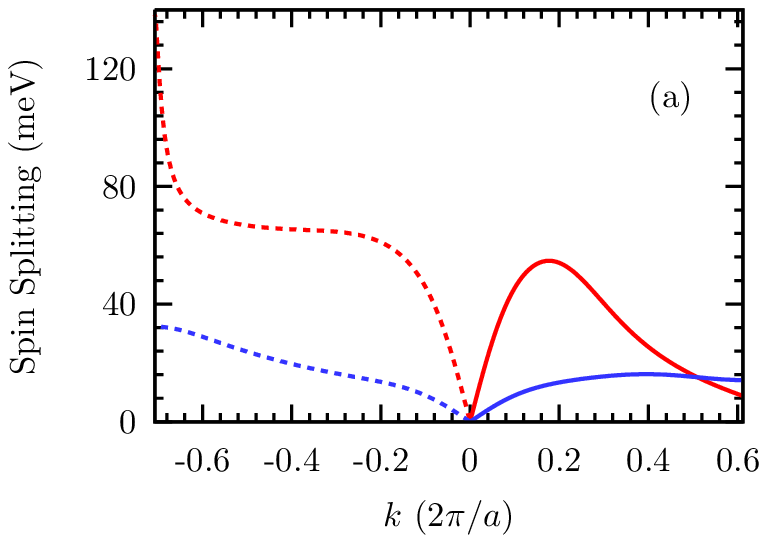,width=6.5cm}\\
 \vskip1.2pc
 \epsfig{file=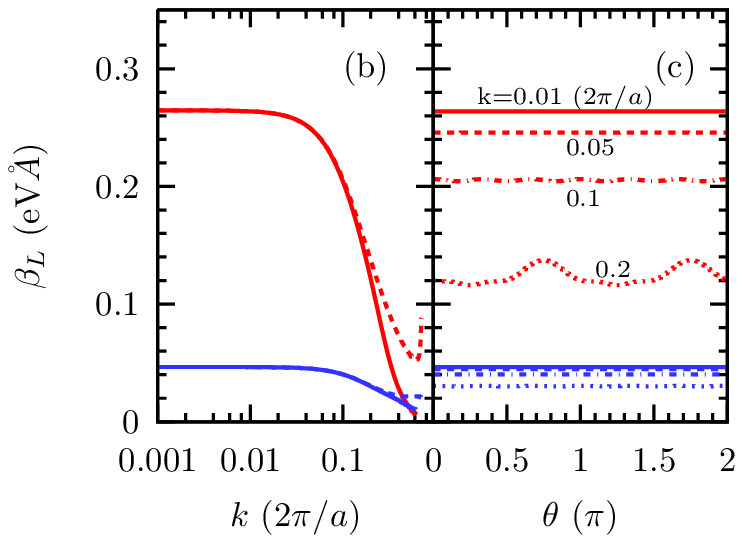,width=6.5cm}
  \caption{(Color Online) (a) Spin splitting in $L$-valley for
    different momentums
    along $L$-$K$ (solid curves) and $L$-$W$ (dashed curves)
    directions;
 (b) The corresponding SOC coefficients $\beta_L$ {\it
      vs.} momentum along $L$-$K$ (solid curves) and $L$-$W$
    (dashed curves) directions;
    (c) $\beta_L$ {\it vs.} angle of the momentum with different
 amplitudes. Solid curves: $k=0.01$; Dashed curves: $k=0.05$; Chain curves: $k=0.1$;
    Dotted curves: $k=0.2$\ $(2\pi/a)$, respectively.
Red/Blue curves are the results from
the $sp^3s^{\ast}d^5$/$sp^3s^{\ast}$ model.
  }
  \label{fig:L}
\end{figure}

\begin{figure}[htb]
  \centering
\epsfig{file=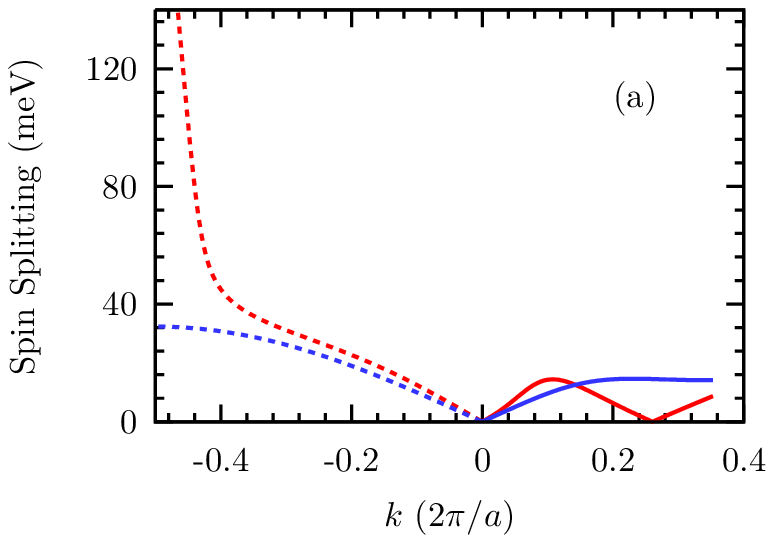,width=6.5cm}\\
\vskip1.2pc
  \epsfig{file=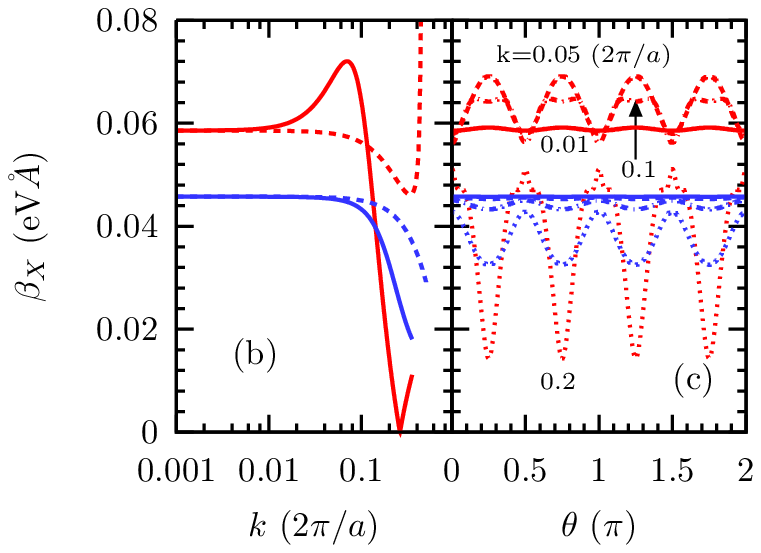,width=6.5cm}
  \caption{(Color Online) (a) Spin splitting in $X$-valley
    for different momentums
    along $X$-$K$ (solid curves) and $X$-$W$ (dashed curves)
    directions;
    (b) The corresponding SOC coefficient $\beta_X$ {\it
      vs.} momentum along $X$-$K$ (solid curves) and $X$-$W$
    (dashed curves) directions;
    (c) $\beta_X$ {\it vs.} angle of the momentum with different
amplitudes. Solid curves: $k=0.01$; Dashed curves: $k=0.05$; Chain curves: $k=0.1$;
Dotted curves: $k=0.2$\ $(2\pi/a)$, respectively.
Red/Blue curves are the results from
the $sp^3s^{\ast}d^5$/$sp^3s^{\ast}$ model.
  }
  \label{fig:X}
\end{figure}

We now turn to the states in $L$- and $X$-valleys.
The spin splittings for different momentums are plotted in
Figs.\ \ref{fig:L} and \ref{fig:X}(a) respectively.
One can see from the figures that, for the same amount of momentum
variation from the valley bottom, the spin splitting in $L$/$X$-valley is much
larger than that in $\Gamma$-valley.
For the states near the $X$-point, the SOC is in the form \cite{ivchenko},
$\beta_X\mathbf{\Omega}_{X}(\mathbf{k})\cdot\mbox{\boldmath${\sigma}$\unboldmath}$ with
\begin{equation}
  \label{eq:3}
 \beta_X \mathbf{\Omega}_X(\mathbf{k})
  =\beta_X(k_x,-k_y,0)\ .
\end{equation}
Near the $L$-valley bottom,
the SOC  reads $\beta_L\mathbf{\Omega}_{L}(\mathbf{k})\cdot
\mbox{\boldmath$\sigma$\unboldmath}$. In the
new coordinate system spanned by [1$\bar{1}$0] ($x^\prime$-axis), [11$\bar{2}$]
($y^\prime$-axis) and [111] ($z^\prime$-axis) vectors,
$\mathbf{\Omega}_L(\mathbf{k})$ reads \cite{ivchenko}
\begin{equation}
  \label{eq:2}
 \beta_L \mathbf{\Omega}_L(\mathbf{k})
  =\beta_L(k_y^\prime,-k_x^\prime,0)\ .
\end{equation}
 In the
above two equations, $\mathbf{k}$ represents the momentum
vector measured from $L$/$X$-valley bottom. These couplings
give the splitting linear to the first order of momentum around the
valley bottoms. The corresponding splitting coefficients
$\beta_{L/X}=\Delta{E}(k)/2k$ for different momentum are plotted in
Figs.\ \ref{fig:L}(b) and \ref{fig:X}(b) respectively. Similar to
that of $\Gamma$-valley, these coefficients are constants near the
valley bottoms. In the $X$-valley, the values of $\beta_X$ obtained from
$sp^3s^{\ast}d^5$ and $sp^3s^{\ast}$ models are close to each other,
i.e., $\beta_X=0.059$ and 0.046\ eV$\cdot$\AA, respectively.
 However, in the $L$-valley, $\beta_L$ determined
from these two models are quite different. For $sp^3s^{\ast}d^5$,
$\beta_L=0.26$\ eV$\cdot$\AA; while for $sp^3s^{\ast}$,
$\beta_L=0.047$\ eV$\cdot$\AA.
This profound difference implies that the $d$ orbit plays an important
role in the spin splitting in the $L$-valley of the lowest conduction
band. This is because the symmetry imposes a $d$-orbital component
in the $L$-valley and $sp^3s^{\ast}d^5$ model can
account this symmetry more accurate. It has been revealed that the
inclusion of the $d$ orbit greatly improves the accuracy of the effective
mass in $L$-valley \cite{Jancu2,Singh,Chang,Richardson1,Boguslawsky}.
Therefore, in our opinion
the spin splitting determined by $sp^3s^{\ast}d^5$ model in $L$-valley
is also more reliable than that by $sp^3s^{\ast}$ model.

\section{Conclusion}

In conclusion we study the SOC in the whole Brillouin
zone for GaAs using both $sp^3s^{\ast}d^5$ and $sp^3s^{\ast}$
nearest-neighbor TB models. For the parameter sets we use,
the spin splittings calculated from both models are in good agreement
with experimental data in the $\Gamma$-valley. We then further explicitly
present the coefficients of the spin splitting in the  $L$ and $X$
valleys. These results are useful for understanding the spin dynamics far away from
equilibrium.

\ack

This work was supported by the Natural Science Foundation of China
under Grant Nos.\ 10574120 and 10725417, the National Basic Research
Program of China under Grant No.\ 2006CB922005 and the Knowledge
Innovation Project of Chinese Academy of Sciences. FJY was partially
supported by China Postdoctoral Science Foundation. One of the authors (M.W.W.)
would like to thank  Mikhail Nestoklon at Ioffe Institute
who first provided us the  number of $\beta_L$ at $L$-valley using
$sp^3s^\ast$ model.


\begin{thebibliography}{100}
\expandafter\ifx\csname url\endcsname\relax
  \def\url#1{\texttt{#1}}\fi
\expandafter\ifx\csname urlprefix\endcsname\relax\def\urlprefix{URL }\fi

\bibitem{wolf}
S.~A. Wolf, J. Supercond.: Incorporating Novel Magnetism 13 (2000) 195.

\bibitem{zutic_2004}
I.~\v{Z}uti\'c, J.~Fabian, S.~D. Sarma, Rev. Mod. Phys. 76 (2004) 323.

\bibitem{dp}
M.~I. D'yakonov, V.~I. Perel', Zh. Eksp. Teor. Fiz. 60 (1971) 1954, [Sov.
  Phys.-JETP {\bf 33}, 1053 (1971)].

\bibitem{dpb}
M.~I. D'yakonov, V.~I. Perel', Fiz. Tverd. Tela 13 (1971) 3581, [Sov. Phys.
  Solid State {\bf 13}, 3023 (1972)].

\bibitem{ras}
Y.~A. Bychkov, E.~I. Rashba, J. Phys. C 17 (1984) 6039.

\bibitem{rashba}
Y.~A. Bychkov, E.~I. Rashba, Pis'ma Zh. Eksp. Teor. Fiz. 39 (1984) 66.

\bibitem{Cardona}
M.~Cardona, N.~E. Christensen, G.~Fasol, Phys. Rev. B 38 (1988) 1806.

\bibitem{Jancu1}
J.-M. Jancu, R.~Scholz, G.~C.~L. Rocca, E.~A. de~Andrada~e Silva, P.~Voisin,
  Phys. Rev. B 70 (2004) 121306(R).

\bibitem{Jancu2}
J.-M. Jancu, R.~Scholz, F.~Beltram, F.~Bassani, Phys. Rev. B 57 (1998)
  6493.

\bibitem{ivchenko}
E.~L. Ivchenko, G.~E. Pikus, Superlattices and Other Heterostructures,
  Springer, Berlin, 1995.

\bibitem{saikin_06}
S.~Saikin, M.~Shen, M.-C. Cheng, J. Phys.: Condens. Matter 18 (2006) 1535.

\bibitem{shen_2004}
M.~Shen, S.~Saikin, M.-C. Cheng, V.~Privman, Mathematics and Computers in
  Simulation 65 (2004) 351.

\bibitem{Vogl}
P.~Vogl, H.~P. Hjalmarson, J.~D. Dow, J. Phys. Chem. Solids 44 (1983) 365.

\bibitem{Santos}
P.~V. Santos, M.~Willatzen, M.~Cardona, A.~Cantarero, Phys. Rev. B 51 (1995)
  5121.

\bibitem{Klimeck}
J.~Klimeck, R.~C. Bowen, T.~B. Boykin, T.~A. Cwik, Superlattices and
  Microstructures 27 (2000) 519.

\bibitem{Boykin1}
T.~B. Boykin, G.~Klimeck, R.~C. Bowen, R.~Lake, Phys. Rev. B 56 (1997)
  4102.

\bibitem{Boykin2}
T.~B. Boykin, G.~Klimeck, R.~C. Bowen, F.~Oyafuso, Phys. Rev. B 66 (2002)
  125207.

\bibitem{Diaz}
J.~G. D\"iaz, G.~W. Bryant, Phys. Rev. B 73 (2006) 075329.

\bibitem{krich_07}
J.~J. Krich, B.~I. Halperin, Phys. Rev. Lett. 98 (2007) 226802.

\bibitem{chantis_06}
A.~N. Chantis, M.~van Schilfgaarde, T.~Kotani, Phys. Rev. Lett. 98 (2006)
  086405.

\bibitem{richards_93}
D.~Richards, B.~Jusserand, H.~Peric, B.~Etienne, Phys. Rev. B 47 (1993)
  16028.

\bibitem{jusserand_95}
B.~Jusserand, D.~Richards, G.~Allan, C.~Priester, B.~Etienne, Phys. Rev. B 51
  (1995) 4707.

\bibitem{richards_96}
D.~Richards, B.~Jusserand, G.~Allan, C.~Priester, B.~Etienne, Solid-State
  Electron. 40 (1996) 127.

\bibitem{maruschak_1983}
V.~I. Marushak, T.~V. Lagunova, M.~N. Seepanova, A.~N. Titkov, Fiz. Tverd. Tela
  25 (1983) 2140.

\bibitem{aronov}
A.~G. Aronov, G.~E. Pikus, A.~N. Titkov, Zh. Eksp. Teor. Fiz. 84 (1983) 1170,
  [Sov. Phys.-JETP {\bf 57}, 680 (1983)].

\bibitem{miller_03}
J.~B. Miller, D.~M. Zumb\"uhl, C.~M. Marcus, Y.~B. Lyanda-Geller,
  D.~Goldhaber-Gordon, K.~Campman, A.~C. Gossard, Phys. Rev. Lett. 90 (2003)
  076807.

\bibitem{zhou_prb_2007}
J.~Zhou, J.~L. Cheng, M.~W. Wu, Phys. Rev. B 75 (2007) 045305.


\bibitem{Singh}
S.~B. Singh, C.~A. Singh, Am. J. Phys. 57 (1989) 894.

\bibitem{Chang}
Y.-C. Chang, D.~E. Aspnes, Phys. Rev. B 41 (1990) 12002.

\bibitem{Richardson1}
S.~L. Richardson, M.~L. Cohen, S.~G. Louie, J.~R. Chelikowsky, Phys. Rev. B 33
  (1986) 1177.

\bibitem{Boguslawsky}
P.~Boguslawsky, I.~Gorczyca, Semicond. Sci. Technol. 9 (1994) 2169.

\end{thebibliography}

\end{document}